\documentclass[aps,prl,amsfonts,nofootinbib]{revtex4}

\usepackage{amssymb,amsmath}
\usepackage{graphicx}
\usepackage{epsfig}

\newcommand{\beq}{\begin{equation}}
\newcommand{\eeq}{\end{equation}}
\newcommand{\bea}{\begin{eqnarray}}
\newcommand{\eea}{\end{eqnarray}}
\newcommand{\vc}[1]{{\textbf{#1}}}
\newcommand{\mc}[1]{\mathcal{#1}}

\begin{document}
 
\title{Thermal Interpretation of Infrared Dynamics in de Sitter}

\author{Gerasimos Rigopoulos}
\affiliation{School of Mathematics and Statistics, 
	Newcastle University, 
	Newcastle upon Tyne, NE1 7RU, UK}

\begin{abstract}
\noindent 
The infrared dynamics of a light, minimally coupled scalar field in de Sitter spacetime with Ricci curvature $R=12H$, averaged over horizon sized regions of physical volume $V_H=\frac{4\pi}{3}\left(\frac{1}{H}\right)^3$, can be interpreted as Brownian motion in a medium with de Sitter temperature $T_{DS}=\frac{\hbar H}{2\pi}$. We demonstrate this by directly deriving the effective action of scalar field fluctuations with wavelengths larger than the de Sitter curvature radius and generalizing Starobinsky's seminal results on stochastic inflation. The effective action describes stochastic dynamics and the fluctuating force drives the field to an equilibrium characterized by a thermal Gibbs distribution at temperature $T_{DS}$ which corresponds to a de Sitter invariant state. Hence, approach towards this state can be interpreted as thermalization. We show that the stochastic kinetic energy of the coarse-grained description corresponds to the norm of $\partial_\mu\phi$ and takes a well defined value per horizon volume $\frac{1}{2}\langle \left(\nabla\phi\right)^2\rangle = - \frac{1}{2}T_{DS}/V_H$. This approach allows for the non-perturbative computation of the de Sitter invariant stress energy tensor $\langle T_{\mu\nu} \rangle$ for an arbitrary scalar potential.   
\end{abstract}

\maketitle

\section{Introduction}

De Sitter spacetime is one of the most fundamental solutions to the equations of General Relativity with particular physical relevance: it's the archetypal inflationary spacetime describing the primordial universe, while the present universe is also driven into a similar de Sitter or quasi-de Sitter phase. Its classical geometrical properties have been known for a long time and the study of quantum fields in a de Sitter background has been extensively pursued. Nevertheless, a clear cut picture of quantum effects in de Sitter has not been reached. The de Sitter horizon is associated with a temperature\footnote{We set the speed of light $c=1$ and the Boltzmann constant $k_B=1$, but keep $\hbar$ to make the quantum mechanical nature evident.}                       
\beq\label{DS-Temp}
T_{DS}=\frac{\hbar H}{2\pi}\,,
\eeq
a physical manifestation of which is the thermal spectrum of excitations exhibited by an Unruh-De Witt detector coupled to a scalar field in de Sitter, making de Sitter act as a heat bath for an observer confined within the horizon \cite{Gibbons:1977mu, Birrell:1982ix}. However such notion is not without paradoxes and the effects of this temperature are still not entirely clarified \cite{Spradlin:2001pw}. On the other hand, massless/light scalar fields on \emph{superhorizon} scales also exhibit fluctuations. These fluctuations are highly relevant for inflation \cite{Mukhanov:2005sc} but have not been demonstrated to exhibit a direct thermodynamical link to the temperature (\ref{DS-Temp}); although typical fluctuations on large scales are of order $T$, inflationary fluctuations have a scale invariant and not a thermal spectrum. In this letter we provide a thermal interpretation of the field's IR dynamics.           

Let us describe de Sitter spacetime using flat slicing coordinates in which the metric reads
\beq\label{flat}
ds^2=-dt^2+e^{2Ht}d\vc{x}^2\,,
\eeq
and assume that the de Sitter phase starts at some definite time, labeled by $t=0$. The dynamics of a test scalar field is governed by the action
\bea\label{Model}
S=\int d^4x \, a^3 \Big[\frac{1}{2}\dot{\Phi}^2 - \frac{1}{2} \frac{\left(\partial_{i}\Phi\right)^2 }{a^2} - U(\Phi) \Big]\,,
\eea
leading to the classical Klein-Gordon equation
\beq\label{KG1}
\ddot{\phi} + 3H\dot{\phi}-a^{-2}\nabla^2_{\vc{x}}\,\phi+\frac{\partial U}{\partial\phi}=0\,,
\eeq
where $a=e^{Ht}$ is the scale factor. We will consider a light scalar for which $0\leq m\ll \hbar H$ and focus on long wavelengths $r>1/H$ ($r=e^{Ht}x$) where the spatial gradients do not influence the dynamics. If $3H$ is interpreted as a ``friction'' term, one would also expect the presence of an associated fluctuating force      
\beq\label{KG2}
\ddot{\phi}+3H\dot{\phi}+\frac{d U}{d\phi}=\xi(t)
\eeq
with the amplitude of the fluctuation given by 
\beq\label{KG-noise}
\langle\xi(t)\xi(t')\rangle = \frac{(2\times 3H \times T_{DS})}{\frac{4\pi}{3}\left(\frac{1}{H}\right)^3}\,\delta(t-t')\,.
\eeq
The inverse volume factor is required for dimensional reasons; we choose it to simply be the physical volume of the de Sitter horizon $R_{\rm DS}=1/H$. This relation mirrors known fluctuation-dissipation relations in thermal systems.

In what follows we show from first principles that (\ref{KG2}) and (\ref{KG-noise}) are correct, using functional integral methods. In particular, we demonstrate that the field $\phi$, when averaged over regions of volume $\frac{4\pi}{3}R^3_{\rm DS}$, executes Brownian motion as if coupled to an environment at temperature $T_{\rm DS}$.\footnote{For a recent study of the Brownian motion of a particle coupled to a scalar field inside the horizon see \cite{Fischler:2014tka}.} Such a stochastic force will drive the field from any initial state with Bunch-Davies short-scale behaviour to a thermal Gibbs equilibrium (see (\ref{MB-dist})) on large wavelengths and at late times, and in this sense the existence of a stochastic force satisfying (\ref{KG-noise}) \emph{defines} the system as thermal. This thermal equilibrium corresponds to a de Sitter invariant state and approaching it can be interpreted as thermalization at temperature $T_{DS}$. This appears in line with cosmic no-hair theorems discussed in \cite{Hollands:2010pr, Marolf:2010nz, Marolf:2011sh} although these works do not discuss a thermal interpretation.

The idea that the dynamics of long wavelength light fields in (quasi-)de Sitter can be described by a stochastic Langevin equation was first explicitly proposed by Starobinsky \cite{Starobinsky:1986fx} (see also \cite{Starobinsky:1994bd}) and was based on the over-damped, first order version of (\ref{KG2}),
\beq\label{starob1}
\dot{\phi} +\frac{d U/d\phi}{3H} = \tilde{\xi}(t)\,,\quad 
\langle\tilde{\xi}(t)\tilde{\xi}(t')\rangle = \frac{\hbar H^3}{4\pi^2}\,\delta(t-t')\,.
\eeq
The treatment presented here extends Starobinsky's original stochastic inflation approach, to the full second order dynamics. The correspondence of the predictions of (\ref{starob1}) for field correlators $\langle\phi^n\rangle$ with those of Quantum Field Theory at the perturbation level was first pointed out in \cite{Woodard:2005cw,Tsamis:2005hd} and later elucidated further in \cite{vanderMeulen:2007ah, Finelli:2008zg,Garbrecht:2013coa, Garbrecht:2014dca,Onemli:2015pma}. Below, we exhibit this link from first principles by showing that the effective long wavelength action of a light scalar in de Sitter is equivalent to that of a particle in thermal Brownian motion. Finally, our approach includes kinetic terms and provides a non-pertubative expression for the de Sitter invariant stress-energy tensor 
\beq
\left\langle T_{\mu\nu}\right\rangle = -g_{\mu\nu} \left(\left\langle U \right\rangle - \frac{1}{4}\left\langle \phi\frac{dU}{d\phi} \right\rangle\right)\,,
\eeq
where brackets denote averaging over the equilibrium distribution.   
        
\section{Stochastic dynamics in de Sitter}

The appropriate framework for describing the real time evolution of the scalar field from given initial conditions, without reference to a future asymptotic state, is the so-called Schwinger-Keldysh formalism employing the amphichronous or closed-time-path path integral \cite{Calzetta:2008iqa, Altland:2006si}
\beq   
Z\left[\vc{J}\right] = \int D\boldsymbol{\phi}\, \exp \frac{i}{2\hbar} \int_x\left( \boldsymbol{\phi}^T\mathbf{D}^{-1}\boldsymbol{\phi} -2\tilde{V}(\phi)+\boldsymbol{\phi}^T\vc{J}\right)\,, 
\eeq
where $\int_x = \int d^4x\sqrt{-g}\,,$ and
\beq
\boldsymbol{\phi}=\left(\begin{matrix}
\phi_1\\\phi_2	
\end{matrix}\right)\,,\quad \vc{J}=\left(\begin{matrix}
J_1\\-J_2	
\end{matrix}\right)\,,\quad \tilde{V}(\boldsymbol{\phi})=V(\phi_1)-V(\phi_2)\,,\quad \mathbf{D}^{-1}(x,x')=\left(\begin{matrix}
\left(\nabla^2-m^2\right) & 0\\ 0 & -\left(\nabla^2-m^2\right)	
\end{matrix}\right)\delta(x-x')\,.
\eeq
The boundary conditions assume some initial state or density matrix in the past and that $\phi_+=\phi_-$ at some point in the future after any possible time of interest. This determines the way the differential operator is to be inverted:
\beq
\left(\begin{matrix}
	\left(\nabla^2-m^2\right) & 0\\ 0 & -\left(\nabla^2-m^2\right)	
\end{matrix}\right)\,\mathbf{D}(x,x')=
\frac{\delta(x,x')}{\sqrt{-g}}\,,
\eeq
defining the propagator in the $\phi_{1,2}$ basis
\beq
\mathbf{D}(x,x')=\left(\begin{matrix}
	D_{11}(x,x')&	D_{12}(x,x')\\	D_{21}(x,x') &	D_{22}(x,x')	
\end{matrix}\right)= -i\left(\begin{matrix}
\langle T \phi(x)\phi(x')\rangle & \langle\phi(x)\phi(x')\rangle\\\langle\phi(x')\phi(x)\rangle & \langle\tilde{T}\phi(x)\phi(x')\rangle	
\end{matrix}\right)\,.
\eeq
A more physical description employs the ``classical'' and ``quantum'' fields 
\beq
\phi=\frac{\phi_1+\phi_2}{2}\,,\quad \phi_q=\phi_1-\phi_2\,,
\eeq
and in this basis the propagator takes its Keldysh form 
\bea
\mathbf{D} &=& \mathbf{U}\left(\begin{matrix}
	D_{11}(x,x')&	D_{12}(x,x')\\	D_{21}(x,x') &	D_{22}(x,x')	
\end{matrix}\right)\mathbf{U}^{\rm T} = \left(\begin{matrix}
-i F(x,x')&	D_{\rm R}(x,x')\\	D_{\rm A}(x,x') & 0	
\end{matrix}\right) \\
&=& -i \left(\begin{matrix}
\frac{1}{2}\langle \{\phi(x),\phi(x')\} \rangle & \langle[\phi(x),\phi(x')]\rangle \Theta(t-t')\\ \langle[\phi(x'),\phi(x)]\rangle \Theta(t'-t)& 0
\end{matrix}\right)\,,
\eea
where
\beq
\mathbf{U}=\left(\begin{matrix}
	\frac{1}{2}& 	\frac{1}{2} \\1 & 1
\end{matrix}\right)
\eeq
is the transformation matrix between the two bases. $D_{R(A)}$ is the retarded (advanced) Green function and $F$ is the Keldysh component of the propagator. Furthermore, in this basis
\beq
\mathbf{D}^{-1}(x,x')= \left(\begin{matrix}
	0& (\nabla^2-m^2) \\ (\nabla^2-m^2) & 0	
\end{matrix}\right)\delta(x-x')\,,
\eeq 
\beq
\tilde{V} =\frac{\partial V}{\partial \phi_c}\phi_q + \sum\limits_{m=1}^\infty\!\frac{V^{{(2m+1)}}(\phi_c)}{\left(2m+1\right)!}\left(\frac{\phi_q}{2}\right)^{2m+1}\,.
\eeq
       
Let us now derive the effective action for the long wavelength part of the system. A Gaussian-like integral satisfies 
\beq
\int\limits_{-\infty}^{+\infty}  dx \, \exp\left({-\frac{x^2}{2\alpha} - V(x)}\right) = \sqrt{\frac{2\pi\beta \gamma}{\alpha}}\int  \limits_{-\infty}^{+\infty}  dydz\, \exp \left({-\frac{y^2}{2\beta} -\frac{z^2}{2\gamma} - V(y+z)} \right) \,,
\eeq
where
$x=y+z$ and $\alpha = \beta + \gamma $.
We can promote this relation to the functional integral, ignoring the field independent factor, by splitting the propagator 
\beq\label{split}
\mathbf{D}(x,x')=\mathbf{D}_>(x,x')+\mathbf{D}_<(x,x')\,.
\eeq     
To achieve the split (\ref{split}) we use a window function $W(x,x')$, smoothing out short wavelength perturbations, as well as its complementary window function $\bar{W}(x,x')$, filtering out long wavelength fluctuations, satisfying  
\beq\label{sum windows}
W(x,x') + \bar{W}(x,x') = \delta(x,x') \,.
\eeq
Here, long and short are defined with respect to some smoothing scale which could be both spatial and temporal. We normalize $\int_{x'} W(x,x') =1$ so that $\int_{x'}\bar{W}(x,x')=0$. We take the $\mathbf{D}_<$ propagator to be 
\beq\label{long} 
\mathbf{D}_<(x,x') = \int_y \int_z W(x,y)\mathbf{D}(y,z)W(z,x') \equiv W\mathbf{D}W\,,
\eeq
and therefore from (\ref{sum windows}) $\mathbf{D}_>$ is given by
 \beq\label{short}
 \mathbf{D}_>=\bar{W}\mathbf{D}\bar{W} + W\mathbf{D}\bar{W} + \bar{W}\mathbf{D}W\,.
 \eeq
Note that apart from the purely short wavelength component, $\mathbf{D}_>$ contains cross-terms involving both $W$ and $\bar{W}$. These terms are only relevant around the smoothing scale and will tend to zero as the window functions become sharp in Fourier space. For any reasonable window function, $D_>$ and $D_<$ will not overlap for scales sufficiently different from the smoothing scale. The field integral can then be written as 
\beq\label{Z}
Z[\vc{J}]=\int D\boldsymbol{\phi}_< D\boldsymbol{\phi}_> \exp \frac{i}{2\hbar}\left[{ \boldsymbol{\phi}_>^T \mathbf{D}_>^{-1} \boldsymbol{\phi}_> + \boldsymbol{\phi}^T_< \mathbf{D}_<^{-1} \boldsymbol{\phi}_< - \tilde{V}(\boldsymbol{\phi}_{<}+\boldsymbol{\phi}_{>})}+\left(\boldsymbol{\phi}^T_<+\boldsymbol{\phi}^T_>\right)\vc{J}\right]\,,
\eeq
where the full field $\phi$ has been split as
\beq
\phi=\phi_> + \phi_< \,, 
\eeq
with the fluctuations of $\phi_<$ and $\phi_>$ governed by the propagators $D_<$ and $D_>$ respectively.  

Let us emphasize here that using a smooth window function $W$ does not partition the function space in which the field $\phi$ lives into fields with strictly short and strictly long modes. An exact step function in $k$-space, and only such a projector function, would be required for this. Therefore, the `integration variables' $\phi_<$ and $\phi_>$, integrated over in the path integral (\ref{Z}), contain all wavelengths in their integration measure. However, by construction $D_<$ is suppressed on short scales and hence configurations of $\phi_<$ with short wavelength components have suppressed contributions to the path integral due to a correspondingly large exponent. Similarly, the propagator (\ref{short}) ensures that long wavelength configurations do not contribute in the $D\phi_>$ path integral. In this sense $\phi_<$ and $\phi_>$ can meaningfully be considered as long and short wavelength fields respectively. Note that this formulation differs from using a direct convolution of $\phi$ with a window function as was originally done in \cite{Starobinsky:1986fx} (see also \cite{Morikawa:1989xz, Levasseur:2013ffa}), bringing this approach more in line with common notions of renormalization, see \cite{Delamotte:2007pf}. More importantly, as we will see the split of the fields into long and short wavelength components is time dependent in inflation. Given this time dependence, the common practice of convolving the field with a window function does not offer itself for a clear understanding of the integration measure's split into long and short wavelength sectors: $D\phi\, \rightarrow \, D\phi_< D\phi_>$.

It is now straightforward to perform the integration over $\phi_>$ obtaining
\beq\label{Z only long}
Z[\vc{J}]=\int D\boldsymbol{\phi}_< \,\, \exp \frac{i}{2\hbar}\left( \boldsymbol{\phi}^T_< \mathbf{D}_<^{-1} \boldsymbol{\phi}_<- \tilde{V}(\boldsymbol{\phi}_<)+\Delta S +\boldsymbol{\phi}^T_<\vc{J}\right)\,,
\eeq
where we have also dropped the term $-\frac{i}{2}\vc{J}^T\mathbf{D}_>\vc{J}$ that arises from the integration. This is always possible if we chose to probe only long wavelegth fields and thus use a source for which $\bar{W}\vc{J}\simeq 0$. The term $\Delta S (\boldsymbol{\phi}_<)$ arises from the non-linear interactions in $V$. We expect that any UV divergences present in $\Delta S$  will resemble those in Minkowski spacetime and will be treatable following the usual procedures. We will investigate this as well as the contribution to the effective IR action in a forthcoming publication \cite{ian-gerasimos}.     

We now need to find the operator inverse of $D_<$. We can formally write
\bea\label{expansion}
\mathbf{D}_<^{-1}=\frac{1}{\mathbf{D}-\mathbf{D}_>}&=&\mathbf{D}^{-1} + \mathbf{D}^{-1}\mathbf{D}_>\mathbf{D}^{-1}+\ldots\nonumber\\
&=&\mathbf{D}^{-1}  -  \mathbf{D}^{-1}\bar{W}\mathbf{D}\bar{W}\mathbf{D}^{-1}
+\mathbf{D}^{-1}\bar{W}+ \bar{W}\mathbf{D}^{-1}+\ldots
\eea   
where in the second line we used (\ref{short}) and the operator series on the r.h.s is understood to act on functions with only long wavelength support. Its truncation to the first two terms will be accurate if $D_<$ is suppressed compared to $D$ on large scales. This does happen in de Sitter space since the propagator of a light, minimally coupled field decays with a very mild power law over large distances due to inflationary infrared enhancement \cite{Garbrecht:2013coa}, unlike $D_>$ which by construction tends to zero. Note that this argument would not hold if the non-smoothed $D$ also tended to zero at large distances, as happens for example in Minkowski space. Thus we find that the dominant contribution to the long wavelength action, expressed in the Keldysh basis, is 
\beq\label{quadratic LW}
\boldsymbol{\phi}^T\mathbf{D}_<^{-1} \boldsymbol{\phi} = \left(\begin{matrix} \phi_c,&\phi_q\end{matrix}\right)\left(\begin{matrix}
	0& (\nabla^2-m^2) \\ (\nabla^2-m^2) & 0	
\end{matrix}\right)\left(\begin{matrix}\phi_c\\ \phi_q
\end{matrix}\right) - \left(\begin{matrix} \phi_c,&\phi_q\end{matrix}\right)\left(\begin{matrix}
0& \nabla^2\bar{W}D_A\bar{W}\nabla^2 \\\nabla^2\bar{W}D_R\bar{W}\nabla^2& -i\nabla^2\bar{W}F\bar{W}\nabla^2	
\end{matrix}\right)\left(\begin{matrix}\phi_c\\ \phi_q
\end{matrix}\right) \,,
\eeq  
where we removed the subscript $<$ to simplify notation and integration by parts is understood on the second matrix. Any terms where $\bar{W}$ is directly convolved with $\phi$ only provide higher order derivative terms which are subdominant and were dropped from (\ref{quadratic LW}). The $F$ term in this equation corresponds to a \emph{stochastic force} $\xi(x)$ \cite{Martin:1973zz, Altland:2006si} with correlation 
\beq
\langle\xi(x)\xi(x')\rangle \equiv\mc{N}({x},{x}')= \int_{u,v}\nabla^2_{x}\nabla^2_{x'}\bar{W}(x,u)F(u,v)\bar{W}(v,x')\,.
\eeq

The above demonstration of the development of a semi-classical stochastic component in the IR dynamics would be valid for any quantum field theory in which the IR parts of propagators exhibit enhancement relative to the UV parts. A suitably chosen window function $W$ serves to separate the corresponding IR-enhanced from the UV regime. A light scalar field in de Sitter and inflationary quasi-de Sitter spacetimes provides a concrete example for such a system. Note that in this case $D_A$ and $D_R$ are not IR enhanced compared to the Keldysh propagator $F$ which is and provides the dominant contribution to the second term of (\ref{quadratic LW}). Hence the IR dynamics of a light scalar field in these spacetimes is stochastic \cite{Starobinsky:1986fx}. 

Let us now adopt the coordinates (\ref{flat}) for which  
\beq
\nabla^2 \rightarrow \partial_t^2 + 3H \partial_t -{\frac{1}{a^2}{\nabla}_{{\vc{x}}}^2 }\,.
\eeq
To see which terms in the effective action are the relevant operators in the IR let us define a new dimensionless spatial coordinate $\tilde{\vc{x}}=\vc{x}(\epsilon H)^3$ and rescale ${\phi_q}=-\psi a^{-3}{\left(\epsilon H\right)^3\hbar}$. The exponent in the path integral is then written as 
\bea\label{MSRJD-1}
i\frac{S[\phi,\psi]}{\hbar}=&&i\int dt d^3\tilde{\vc{x}} \,\Bigg[\frac{1}{2}\left(\begin{matrix} \phi,&\psi\end{matrix}\right)\left(\begin{matrix}
	0& (-\hat{\nabla}^2+{m}^2) \\ (-\tilde{\nabla}^2+m^2) & -i\hbar\int_{\tilde{x}'}\mc{N}(\tilde{x},\tilde{x}')	
\end{matrix}\right)\left(\begin{matrix}\phi\\ \psi
\end{matrix}\right)+\frac{\partial V}{\partial\phi}\psi\nonumber\\ &&+2\!\sum\limits_{m=1}^\infty\!\frac{V^{{(2m+1)}}}{\left(2m+1\right)!}\left(-\frac{\psi}{2}\right)^{2m+1}\!\left( \frac{\epsilon H}{a}\right)^{6m}\hbar^{2m+1}\Bigg]
\eea
where 
\bea
-\tilde{\nabla}^2 &=& \partial_t^2 + 3H \partial_t -{\frac{H^2\epsilon^2}{a^2}{\nabla}_{\tilde{\vc{x}}}^2 }\\
-\hat{\nabla}^2 &=& \partial_t^2 - 3H \partial_t - {\frac{H^2\epsilon^2}{a^2}{\nabla}_{\tilde{\vc{x}}}^2 }\,.
\eea

The $\epsilon\rightarrow 0$ limit corresponds to increasing coarse-gaining in units of $1/H$. It is clear form the above expression that this scaling also recovers the semi-classical limit by suppressing the $\psi^{2m+1}$ terms and leaving the classical equations of motion along with the stochastic fluctuation term proportional to $\psi^2$, which is the leading quantum effect in the IR. The scaling also suppresses the spatial gradient term compared to the other operators in the action. Thus, on long wavelengths 
\beq\label{MSRJD-2}
i\frac{S[\phi,\psi]}{\hbar}\simeq i\int dt\, d^3\tilde{\vc{x}}\left[\psi\left(\ddot{\phi} +3H\dot\phi+m^2\phi+\frac{\partial V}{\partial\phi}\right)\right]  
-\frac{\hbar}{2}\int d^3\tilde{\vc{x}} d^3\tilde{\vc{x}}'\,dtdt' \,{\psi}(\tilde{x})\mc{N}(\tilde{x},\tilde{x}'){\psi}(\tilde{x}')\,,
\eeq
which describes stochastic Langevin dynamics for $\phi$. Note that the absorbtion of the $a^3$ proper volume factor into $\psi$, and the consequent appearance of a $(-)$ sign in the $\dot{\phi}$ term of $-\hat{\nabla}^2$, allows us to treat this system as experiencing friction determined by the $3H$ coefficient.           

The noise kernel $\mc{N}(\tilde{x},\tilde{x}')$ depends of course on the window function but any physical results should be independednt of this choice. The original formulation of stochastic inflation by Starobinsky, which neglected the field acceleration $\ddot{\phi}$, used a sharp step function in $k$ space to define the long wavelehgth system. Here we use 
\beq
W(t,t',\vc{x},\vc{y})= \delta(t-t')\int \frac{d^3k}{\left(2\pi\right)^3} W_k(t)e^{i\vc{k}\left(\vc{x}-\vc{y}\right)}
\eeq
with
\beq\label{window}
W_k(t)=\left(1-\frac{k^3}{(\epsilon a H)^3}\right)\,\Theta \left[\ln\left(\frac{\epsilon a H}{k}\right)\right]\,,
\eeq
for which 
\beq\label{window 2}
\ddot{W}_k(t)+3H\dot{W}_k(t)=3H \delta(t-\frac{1}{H}\ln\left(k/\epsilon H\right)). 
\eeq
Using that for a massive field in the Bunch Davies vacuum
\beq
F(k,t,t') \simeq \frac{H^2}{2k^3} \left(\frac{k}{a(t)H}\right)^{\frac{m^2}{3H^2}}\left(\frac{k}{a(t')H}\right)^{\frac{m^2}{3H^2}}
\eeq
on long wavelengths \cite{vanderMeulen:2007ah}, we find
\beq\label{N}
\mc{N}(x,x') = \frac{9\hbar H^5}{4\pi^2}e^{\frac{2m^2}{3H^2}\ln \epsilon} \frac{\sin\left(a \left|\tilde{\vc{x}}-\tilde{\vc{x}}'\right|\right)}{a \left|\tilde{\vc{x}}-\tilde{\vc{x}}'\right|}\delta(t-t')\,.
\eeq
The spatial dependence and the white noise property of the noise correlator are directly related to the use of a window function satisfying (\ref{window 2}), which necessarily contains a Heaviside function. Smoother window functions will give correlators with a universal $|\vc{x}-\vc{x}'|^{-4} e^{-2H\Delta t}$ asymptotic profile \cite{Winitzki:1999ve}. We will investigate such more general window functions elsewhere. Note that we as long as we require $|\ln \epsilon|\ll\frac{3H^2}{2m^2}$, (\ref{N}) coincides with the $m=0$ case. In practise this is a very weak constraint and we can always set  $e^{\frac{2m^2}{3H^2}\ln \epsilon}\simeq 1$ 

When inserted in the functional integral and used for \emph{perturbative} calculations, the action (\ref{MSRJD-2}) determines the free correlation functions as
\beq\label{propagators}
\left(\begin{matrix}\langle\phi(t)\phi(t')\rangle & \langle\phi(t)\psi(t')\rangle \\ \langle\psi(t)\phi(t')\rangle & \langle\psi(t)\psi(t')\rangle\end{matrix}\right)=
\left(\begin{matrix}
\mc{F}(t,t')&i\mc{G}^R(t,t')\\
i\mc{G}^A(t,t')&0\end{matrix}\right)\,
\eeq
where $\mc{G}^{(R,A)}(t,t')$ are the retarded and advanced Green functions for $\partial_t^2+3H\partial_t+m^2$
\beq
\mc{G}^R(t,t')=\mc{G}^A(t',t)=\frac{1}{3H}
\left(e^{-\frac{m^2}{3H}(t-t')}-e^{-3H(t-t')}\right)\Theta(t-t')
\eeq
and
\beq\label{corr}
\mc{F}(t,t')=\frac{9\hbar H^5}{4\pi^2}\int\limits_{0}^{+\infty} d\tau \, \mc{G}^R(t,\tau)\mc{G}^A(\tau,t') + \mc{F}_0(t,t')\,,
\eeq
where $\mc{F}_0(t,t')$ satisfies the linear equation of motion and encodes the initial state. It is possible to choose it such that  
\beq\label{corr2}
\langle\phi(t)\phi(t')\rangle = \frac{3\hbar H^4}{8\pi^2m^2}\left(e^{-\frac{m^2}{3H}|t-t'|} -\frac{m^2}{9H^2}e^{-3H|t-t'|} \right)\,,
\eeq
which depends only on $|t-t'|$. Deviations of $F$ from (\ref{corr2}), corresponding to different choices of initial conditions, decay at sufficiently large times $t,t'>H/m^2$ and the leading term in (\ref{corr2}) is recovered, reproducing the standard de Sitter invariant result for $3H|t-t'|>1$ and small spatial separations. Hence the correlator naturally tends to its de Sitter invariant form (see \cite{Marolf:2011sh} for a similar point and also \cite{Markkanen:2016aes}).

The correspondence of the stochastic formalism's Feynman diagram expansion in the over-damped limit (where $\ddot{\phi}$ is ignored) to the pertubative QFT Feynman diagrams in the IR was shown in \cite{Garbrecht:2013coa,Garbrecht:2014dca}. The treatment here has demonstrated the correspondence from first principles: the stochastic formulation is precisely the effective IR theory obtained when sub-Hubble modes are integrated out. Furthermore, equations (\ref{MSRJD-2}) and (\ref{N}) demonstrate that the long wavelength sector of a scalar field in de Sitter can indeed be thought of as a classical system subject to both friction and thermal noise at the de Sitter temperature, related by a classical fluctuation-dissipation relation. This proves the heuristic assertion made in the introduction. As we discuss now, this also implies the existence of a non-pertubative thermal Gibbs equilibrium to which any initial IR state (with Bunch-Davies UV behaviour) eventually relaxes.

\section{Equilibrium distribution and stress energy tensor}
  
To obtain the probability distribution for $\phi$ at a spatial point $\tilde{\vc{x}}$ we consider the quantity
\beq
\mc{P}\left(\phi,t|\phi_i,t_i\right)=\int\limits_{\phi_i}^{\phi} [D\phi] [D\psi] \,\,
e^{i\int^t_{t_i} dt\left[\psi\left(\ddot{\phi} +3H\dot\phi+\frac{\partial U}{\partial\phi}\right)+\frac{i}{2}\frac{9\hbar H^5}{4\pi^2} \psi^2 \right]}
\eeq
where $U=\frac{1}{2}m^2\phi^2+V$, which represents the \emph{probability}\footnote{This is inherited from the closed time path contour in the initial path integral.} to find the field value $\phi$ at time $t$, given the field value $\phi_i$ at $t_{\rm i}$. We can rewrite this as
\beq
\mc{P}\left(\phi,t|\phi_i,t_i\right)=\int\limits_{\phi_i}^{\phi} [D\phi][Dy][D\psi][D\rho] \,\,
e^{\int^t_{t_i} dt\left[i\rho\left(\dot{\phi}-y\right)+i\psi\left(\dot{y} +3Hy+\frac{\partial U}{\partial\phi}\right)-\frac{1}{2}\frac{9\hbar H^5}{4\pi^2} \psi^2 \right]}\,,
\eeq
defining the probability to find both $y$ and $\phi$ as
\beq
\mathcal{W}\left(\phi,y,t|\phi_i,y_i,t_i\right) = \int\limits_{\phi_i,y_i}^{\phi,y} [D\phi][Dy][D\psi][D\rho] \,\,
e^{\int^t_{t_i} dt\left[i\rho\left(\dot{\phi}-y\right)+i\psi\left(\dot{y} +3Hy+\frac{\partial U}{\partial\phi}\right)-\frac{1}{2}\frac{9\hbar H^5}{4\pi^2} \psi^2 \right]}\,.   
\eeq
Writing $\psi=i\psi_E$, $\rho=i\rho_E$
we have
\beq
\mathcal{W}\left(\phi,y,t|\phi_i,y_i,t_i\right) = \int\limits_{\phi_i,y_i}^{\phi,y} [D\phi][Dy][D\psi_E][D\rho_E] \,\,
e^{-\int^t_{t_i} dt\left[\rho_E\dot{\phi}+\psi_E\dot{y}-\mathcal{H}(\psi_E,\rho_E,y,\phi) \right]}
\eeq
which has a canonical structure with the ``pseudo-Hamiltonian'' 
\beq
\mathcal{H}(\psi_E,\rho_E,y,\phi)=\frac{9\hbar H^5}{8\pi^2}\psi_E^2-\psi_E\left(3Hy+\frac{\partial U}{\partial\phi}\right)+\rho_Ey\,.
\eeq
With $\psi_E=-\partial_y$ and $\rho_E=-\partial_\phi$ and normal ordering in the pseudo-Hamiltonian, the probability $\mc{W}$ will satisfy a corresponding ``Schr\"{o}dinger'' equation
which is nothing but the Fokker-Planck equation.
\beq\label{FP1}
\partial_t\mathcal{W}=\left(\frac{9\hbar H^5}{8\pi^2}\frac{\partial^2}{\partial y^2}+3H\frac{\partial}{\partial y}y+\frac{\partial U}{\partial\phi}\frac{\partial}{\partial y}-y\frac{\partial}{\partial\phi}\right)\mc{W}\,.
\eeq
Interestingly, $\mathcal{W}$ can be identified with the the Wigner distribution which also satisfies (\ref{FP1}) \cite{Buryak:1995tx}.

The correlation functions generated by (\ref{MSRJD-2}) can thus be computed using $\mc{W}$. Equilibrium is described by the stationary solution to (\ref{FP1}) which is easily found to be
\beq\label{MB-dist}
\mc{W}(\phi,y)=N^{-1}{e^{-\frac{8\pi^2}{3\hbar H^4}\left(\frac{y^2}{2}+U\right)}}
\eeq
with
$N=\frac{2\sqrt{\pi}}{\sqrt{3}H^2}\int d\phi \,\,e^{-\frac{8\pi^2}{3\hbar H^4}U}$.
Therefore, any late time correlation function $\langle\mc{O}(\phi,\dot{\phi})\rangle$ \emph{in equilibrium} can then be written, after setting $\mc{O}(\phi,\dot{\phi}) \rightarrow \mc{O}(\phi,y)$
\beq
\langle\mc{O}(\phi,y)\rangle
=\int d\phi d{y} \,\,\mc{O}(\phi,y)\,\,\frac{e^{-\frac{8\pi^2}{3\hbar H^4}(\frac{y^2}{2}+U)}}{N}\,.
\eeq
The system therefore equilibrates to a thermal Gibbs distribution which gives
\beq\label{equipartition}
\left\langle y^2\right\rangle_{\rm eq}=\left\langle\phi\frac{dU}{d\phi}\right\rangle_{\rm eq} = \frac{3\hbar H^4}{8\pi^2}\,.
\eeq 
This in turn implies equipartition $\langle V\rangle = \frac{1}{2}\langle y^2\rangle$ for a free field, as is appropriate for a thermal state.  

The velocity $\langle y^2\rangle$ is not the physical velocity of the field at the spatial point $\tilde{\vc{x}}$. It is rather a coarse grained velocity resulting from taking $\epsilon\rightarrow 0$, and the disappearance of spatial gradient terms is an artefact of this limit.\footnote{The author is indebted to Ian Moss for crucial input regarding the following argument and the implications of de Sitter invariance.} It is possible to proceed without introducing $\epsilon$ and keeping the total spacetime gradient at the cost of extra complication that we will address elsewhere. For the purposes if this work we note that for stochastic averages at a single point 
\beq\label{stoch exp}
\left\langle y^2\right\rangle=\left\langle \dot{\phi}_{\rm stoch}^2\right\rangle= \frac{1}{2}\left(\partial_t^2+3H\partial_t\right)\left\langle\phi^2\right\rangle +\left\langle\phi\frac{dU}{d\phi}\right\rangle \,
\eeq
which is consistent with (\ref{equipartition}) at equilibrium. On the other hand, quantum field theoretic expectation values would give
\beq\label{qft exp}
\left\langle\left(\nabla\phi\right)^2\right\rangle_{\rm QFT}= \frac{1}{2}g^{\mu\nu}\nabla_\mu\nabla_\nu\left\langle \phi^2\right\rangle_{\rm QFT} -\left\langle\phi\frac{dU}{d\phi}\right\rangle_{\rm QFT} \,.
\eeq
Since $\nabla_{\vc{x}} \left\langle \phi^2\right\rangle_{\rm QFT}= 0$ in the coordinates (\ref{flat}) and since our results on the coarse grained effective action imply $\left\langle\phi^n\right\rangle_{\rm QFT}= \left\langle\phi^n\right\rangle$ in the IR\footnote{See \cite{Garbrecht:2014dca} for an explicit demonstration of this relation order by order in the IR in perturbation theory for $\lambda\phi^4$.}, we have
\beq
\left\langle\left(\nabla\phi\right)^2\right\rangle_{\rm QFT}=-\left\langle y^2\right\rangle=- \frac{3\hbar H^4}{8\pi^2}\,.
\eeq 
Therefore, de Sitter invariance, for which $\partial_t \langle\phi^n\rangle=0$ is also true, corresponds to the equilibrium distribution (\ref{MB-dist}) per spatial point with $y^2$ representing the full norm of $\partial_\mu\phi$.    

These results allow us to compute the energy momentum tensor of the fluctuations at equilibrium. de Sitter invariance implies  
\beq
\left\langle\label{dS invariance} \nabla_\mu\phi\nabla_\nu\phi\right\rangle=\frac{1}{4}g_{\mu\nu}\left\langle\left(\nabla\phi\right)^2\right\rangle
\eeq  
and hence
\beq
\left\langle T_{\mu\nu}\right\rangle_{\rm eq} = -g_{\mu\nu} \left(\left\langle U \right\rangle_{\rm eq} - \frac{1}{4}\left\langle \phi\frac{dU}{d\phi} \right\rangle_{\rm eq}\right)\,,
\eeq
This expression reproduces known results in renormalized QFT but now $T_{\mu\nu}$ can be computed \emph{non-pertubatively} using ($\ref{MB-dist}$) and is a finite quantity for any well behaved potential function.

\section{Discussion}

The dynamics of a light scalar field on super-Hubble scales in de Sitter is stochastic and the fluctuating ``force'' is precisely such that the field's probability distribution is driven to a Gibbs equilibrium at temperature $T_{DS}$.  Our results extend earlier stochastic treatments and further demonstrate that $\langle\dot{\phi}^2\rangle_{\rm stochastic} = -\langle \left(\nabla\phi\right)^2\rangle_{\rm QFT}$ which allows for the computation of the full stress energy tensor of the stochastically fluctuating field. The equilibrium state is de Sitter invariant and the field's generic evolution towards it can be thought of as thermalization.  Furthermore, note that since euclidean field theory is known to describe equilibrium statistical mechanics, it is perhaps no accident that the late time Lorentzian equilibrium distribution $\mathcal{P}\propto e^{-\frac{8\pi^2}{3H^4}U}$ is also obtained in Euclidean de Sitter computations \cite{Rajaraman:2010xd,Beneke:2012kn}. Finally, the existence of an equilibrium shows that secular divergences appearing in perturbation theory (see eg \cite{Seery:2010kh}) can be re-summed. A variety of approaches, many using stochastic methods, are converging to this conclusion \cite{Riotto:2008mv,Burgess:2009bs,Garbrecht:2011gu, Boyanovsky:2012nd,Serreau:2013psa,Lazzari:2013boa,Guilleux:2015pma,Burgess:2015ajz} which is strongly supported by the results of this work.

\section{Acknowledgements}
\noindent The author would like to warmly thank Ian Moss for many crucial suggestions regarding this work. Many thanks also to Tommi Markkanen and Arttu Rajantie for very useful comments.

\end{document}